# Compensation based Dictionary Transfer for Similar Multispectral Image Spectral Super-resolution

Xiaolin Han, *Member, IEEE*, Huan Zhang, Lijuan Niu, and Weidong Sun, *Member, IEEE*

*Abstract*—**Utilizing a spectral dictionary learned from a couple of similar-scene multi- and hyperspectral image, it is possible to reconstruct a desired hyperspectral image only with one single multispectral image. However, the differences between the similar scene and the desired hyperspectral image make it difficult to directly apply the spectral dictionary from the training domain to the task domain. To this end, a compensation matrix based dictionary transfer method for the similar-scene multispectral image spectral super-resolution is proposed in this paper, trying to reconstruct a more accurate high spatial resolution hyperspectral image. Specifically, a spectral dictionary transfer scheme is established by using a compensation matrix with similarity constraint, to transfer the spectral dictionary learned in the training domain to the spectral super-resolution domain. Subsequently, the sparse coefficient matrix is optimized under sparse and low-rank constraints. Experimental results on two AVIRIS datasets from different scenes indicate that, the proposed method outperforms other related SOTA methods.**

## I. INTRODUCTION

HIGH spatial resolution hyperspectral (HHS) images are rich in both spatial and spectral information, which are of great significance in target detection, image classification [1]-[2]. However, there is an inherent contradiction between the spatial and spectral resolution in optical remote sensing imaging. In reality, low spatial resolution hyperspectral (HS) image of similar scene and high spatial resolution multispectral (HMS) images are relatively easy to obtain. Leveraging the spectral information provided by the similar-scene HS images, spectral super-resolution is one of the efficient ways to reconstruct the HHS image from a given HMS image, which has attracted extensive attentions.

Utilizing similar-scene HS images a prior, the existing spectral super-resolution methods can be roughly grouped into two categories, *i.e.* dictionary based and deep learning based methods. More specifically, for the first category, utilizing spectral dictionary learns the spectral information from similar-scene multispectral and hyperspectral (MS-HS) image pairs, Han *et al.* [3] introduced an HHS image reconstruction method with sparse constraint. Yi *et al.* [4] used the MS-HS image pairs from other scenes to learn the spectral dictionary and then reconstructed the HHS image under sparse and low-rank constraint. Based on the PCA dictionary, Chang *et al.* [5] put forward an HHS image reconstruction method. Besides, Gao *et al.* [6] achieved spectral super-resolution of HHS images in non-overlapping regions utilizing sparse and low-rank learning.

As for the second category, deep learning based methods focus on leveraging a large number of image pairs to establish nonlinear spectral mappings from multispectral to hyperspectral images through neural networks [7]. Among them, Martinez *et al.* [8] developed a pre-trained Unet convolutional neural network (CNN) for HHS image reconstruction. Fu *et al.* [9] developed CNN based spectral super-resolution method with a spectral response function selection layer and negative sparse constraints. the above CNN based methods require a huge index space, which implies a large number of training samples to prevent overfitting. To deal with this issue, Han *et al.* [10] constructed a nonlinear spectral mapping in the spectral domain, that effectively reduced the index space. Besides, Zheng *et al.* [11] developed a spatial-spectral residual attention network, that extracts the spatial and spectral information through a two-branch structure. Utilizing two subnetworks for spectral and spatial feature extracting, Dian *et al.* [12] suggested a model guided network for hyperspectral images reconstruction. Different from the above data driven spectral super-resolution methods, He *et al.* [13] combined a data driven approach with an optimization algorithm to enhance spectral resolution of multispectral images by a spectral response function-guided CNN approach.

As introduced above, the above methods have not considered the differences between the obtained similar-scene and the desired hyperspectral image. Indeed, the above dictionary or deep learning based methods can directly apply the relationship or dictionary established with MS-HS image pairs to the other HMS image, while the relationship or dictionary is fixed by the pre-trained neural network or dictionary learning process in the training domain, and is hard to fine-tune since there are no corresponding samples in the spectral super-resolution domain. That is to say, although many similar spectra between MS-HS image pairs and HMS image can be used to fine-tune or to optimize the relationship from the MS to the HS image, but in which form and how to embed these

X. Han is with the School of Mechatronical Engineering, Beijing Institute of Technology, Beijing, China (hxl@bit.edu.cn). H. Zhang and W. Sun are with the Department of Electronic Engineering, Tsinghua University, Beijing, China (zhanghuan19@mails.tsinghua.edu.cn; wdsun@tsinghua.edu.cn). L. Niu is with the Chinese Academy of Medical Sciences and Peking Union Medical College, Beijing, 100021, China (e-mail: niulijuan8197@126.com)

This work was supported in part by the Beijing Natural Science Foundation (3254044), the National Natural Science Foundation (82471999, 41971294), Beijing Institute of Technology Research Fund Program for Young Scholars of China. (*Corresponding authors: Xiaolin Han; Weidong Sun*) kinds of new information into the pre-trained model or dictionary, to achieve more accurate performance for the spectral super-resolution domain has not been considered.

To solve the above issues, a compensation matrix based Dictionary Transfer method for the similar-scene multispectral image Spectral super-resolution (termed as DTS) is proposed in this paper, trying to transfer the spectral dictionary learned in the training domain to the spectral super-resolution domain. At first, a spectral dictionary transfer scheme is established which is optimized by compensation matrix with similarity constraint. Then, based on the transferred dictionary, the coefficient matrix is optimized under sparse and low-rank constraints. Finally, the HHS image is obtained by multiplying the transferred spectral dictionary and coefficient matrix. The major contributions are list as follows.

1) A new dictionary transfer scheme for similar-scene multispectral image spectral super-resolution is proposed for the first time.
2) A strategy to transfer the spectral dictionary learned within the training domain to the spectral super-resolution domain is proposed, which is achieved by a compensation matrix with similarity constraint.
3) A closed-form solution of the optimized sparse coefficient matrix is calculated theoretically, under sparse and low-rank constraints.

## II. PROPOSED METHOD

In the similar-scene multi spectral image spectral super-resolution, the HHS image $\mathbf{X} \in \mathbb{R}^{\lambda_x \times N}$ is expected to reconstruct from the multispectral image $\mathbf{Y} \in \mathbb{R}^{\lambda_y \times N}$, utilizing sufficient spectral information provided by similar-scene hyperspectral image $\mathbf{Z} \in \mathbb{R}^{\lambda_x \times m}$. In which, $\lambda_x$ and $\lambda_y$ denote the band number in $\mathbf{X}$ and $\mathbf{Y}$, $N$ and $M$ denote the corresponding pixel number per band. Moreover, the spectral degraded process of $\mathbf{X}$ and $\mathbf{Z}$ can be formulated by:

$$\mathbf{Y} = \mathbf{LX} + \mathbf{N}_m \tag{1}$$
$$\mathbf{Z}_y = \mathbf{LZ} + \mathbf{N}_y \tag{2}$$

where, $\mathbf{L} \in \mathbb{R}^{\lambda_y \times \lambda_x}$ denotes the spectral response function, $\mathbf{N}_m$ and $\mathbf{N}_y$ denote Gaussian noise. Under the sparse representation, the spectral dictionary $\mathbf{D}_s \in \mathbb{R}^{\lambda_x \times K}$ and $\mathbf{D}_t \in \mathbb{R}^{\lambda_x \times K}$ are used to represent $\mathbf{X}$ and $\mathbf{Z}$ as:

$$\mathbf{X} = \mathbf{D}_t \mathbf{A}_x + \mathbf{N}_x \tag{3}$$
$$\mathbf{Z} = \mathbf{D}_s \mathbf{A}_z + \mathbf{N}_z \tag{4}$$

where, $\mathbf{A}_z \in \mathbb{R}^{K \times m}$ and $\mathbf{A}_x \in \mathbb{R}^{K \times N}$ denote the corresponding sparse coefficient, and $\mathbf{N}_x \in \mathbb{R}^{\lambda_x \times N}$ and $\mathbf{N}_z \in \mathbb{R}^{\lambda_x \times m}$ denote the Gaussian noises. Substituting Eqs.(3) and (4) into Eqs.(1) and (2), we can get:

$$\mathbf{Y} = \mathbf{LD}_t \mathbf{A}_x + \mathbf{N}_m^x \tag{5}$$
$$\mathbf{Z}_y = \mathbf{LD}_s \mathbf{A}_z + \mathbf{N}_y^z \tag{6}$$

As can be seen from Eqs.(5) and (6), the spectral super-resolution process needs to consider the differences between the similar-scene hyperspectral image $\mathbf{Z}$ and the desired HHS image $\mathbf{X}$, *i.e.*, it is necessary to construct a dictionary transfer from the spectral dictionary $\mathbf{D}_s$ to $\mathbf{D}_t$.

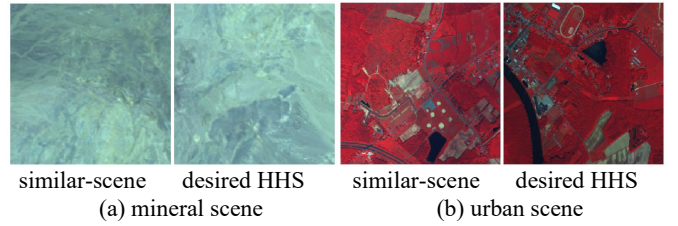

similar-scene    desired HHS      similar-scene    desired HHS
(a) mineral scene         (b) urban scene

Fig.1 False color images of two AVIRIS datasets corresponding to (a) mineral scene and (b) urban scene.

### A. Spectral Dictionary Transfer

If the spectrum $\boldsymbol{z}_i \in \mathbb{R}^{\lambda_x}$ in $\mathbf{Z}$ is closest to the column vector $\boldsymbol{z}_i \in \mathbb{R}^{\lambda_x}$ of the spectral dictionary $\mathbf{D}_s$, it is assumed that the spectrum $\boldsymbol{z}_i$ can be transferred to the spectrum $\boldsymbol{x}_i \in \mathbb{R}^{\lambda_x}$ in the HHS image, though the $i^{th}$ column of compensation matrix $\mathbf{Q}^s \in \mathbb{R}^{\lambda_x \times K}$ as:

$$\boldsymbol{x}_i = \boldsymbol{z}_i + \boldsymbol{q}_i \tag{7}$$

where, $\boldsymbol{q}_i \in \mathbb{R}^{\lambda_x}$ denotes the $i^{th}$ column of $\mathbf{Q}^s$. By left-multiplying the spectral response function $\mathbf{L}$, the compensation matrix $\mathbf{Q}^s$ can be expressed as:

$$\mathbf{Y}^s = \mathbf{Z}_y^s + \mathbf{LQ}^s \tag{8}$$

in which, $\mathbf{Y}^s = [\boldsymbol{y}_1 ... \boldsymbol{y}_i ... \boldsymbol{y}_K]$, $\mathbf{Z}_y^s = \mathbf{LZ}^s = \mathbf{L}[\boldsymbol{z}_1 ... \boldsymbol{z}_i ... \boldsymbol{z}_K]$. Considering the differences in imaging conditions and scenes, the close spectra before and after transfer operation are still similar. Thus, the compensation matrix $\mathbf{Q}^s$ can be optimized by the following minimization problem.

$$\arg\min_{\mathbf{Q}^s} ||\mathbf{Y}^s - \mathbf{Z}_y^s - \mathbf{LQ}^s||_F^2 + \eta ||\mathbf{Q}^s - k\mathbf{Z}^s||_F^2 \tag{9}$$

In Eq.(9), the first and second terms constraint representation error and spectral similarity. Then, the closed-form solutions of Eq.(9) becomes:

$$\mathbf{Q}^s = (\mathbf{L^T L} + \eta \mathbf{I})^{-1}[\mathbf{L^T}(\mathbf{Y}^s - \mathbf{Z}_y^s) + \eta k \mathbf{Z}^s] \tag{10}$$

With the optimized compensation matrix $\mathbf{Q}^s$, the optimization of transferred spectral dictionary $\mathbf{D}_t$ becomes:

$$\arg\min_{\mathbf{D}_t} ||\mathbf{Z}^s + \mathbf{Q}^s - \mathbf{D}_t \mathbf{A}_x^s||_F^2 + \lambda ||\mathbf{A}_x^s||_0 \tag{11}$$

which can be optimized by the widely used K-SVD and OMP algorithm.

### B. Sparse Coefficients Optimization

Considering the spectral similarity mentioned above, the corresponding spectra $\mathbf{Z}^s$ can be represented by the transferred spectral dictionary $\mathbf{D}_t$, expressed by $\mathbf{Z}^s = \mathbf{D}_t \mathbf{A}_u + \mathbf{N}_u$. Thus, represented by the same spectral dictionary $\mathbf{D}_t$, the combination of sparse coefficients $[\mathbf{A}_x \ \mathbf{A}_u]$ has low-rank characteristics. With the optimized compensation matrix $\mathbf{Q}^s$ and transferred spectral dictionary $\mathbf{D}_t$, the optimization of $\mathbf{A}_x$ becomes:

$$\arg\min_{\mathbf{A}_x} ||\mathbf{Y} - \mathbf{LD}_t \mathbf{A}_x||_F^2 + \lambda ||\mathbf{A}_x||_1 + \gamma ||[\mathbf{A}_x \ \mathbf{A}_u]||_* \tag{12}$$

in which, $|| \ ||_1$ and $|| \ ||_*$ denote the convexly approximation of sparse and low-rank constraints. Applying the ADMM [14] algorithm, the above optimization problem can be rewritten as:

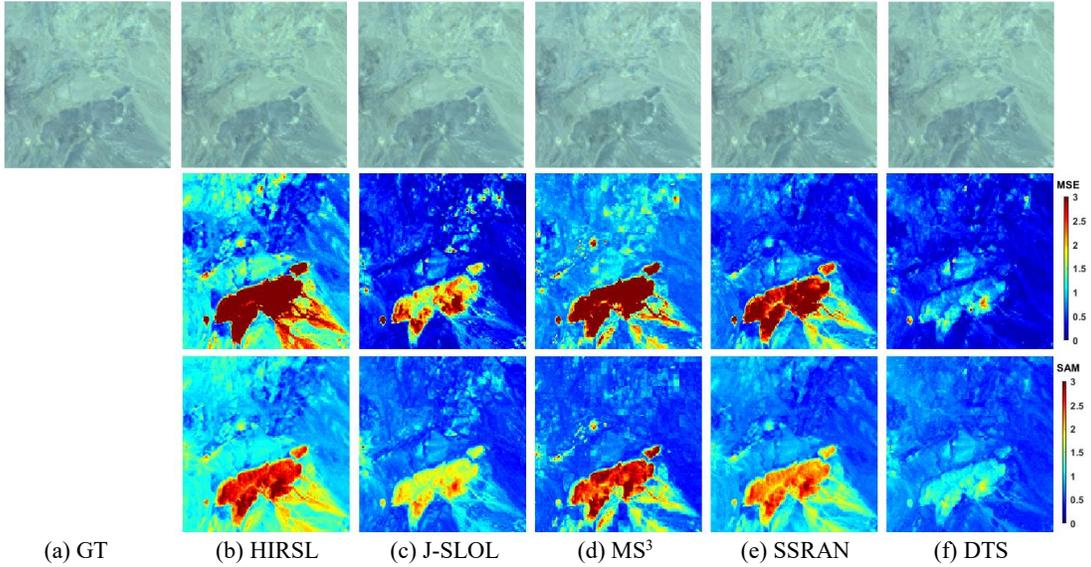

(a) GT      (b) HIRSL      (c) J-SLOL      (d) MS³      (e) SSRAN      (f) DTS

Fig.2 False color image and Error images in MSE and SAM of the mineral scene AVIRIS dataset.

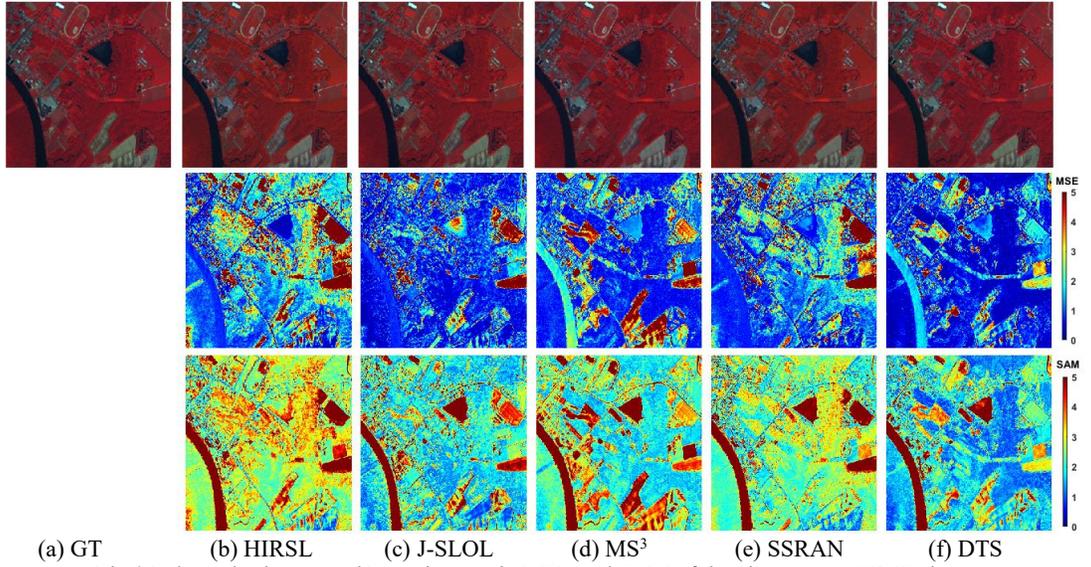

(a) GT      (b) HIRSL      (c) J-SLOL      (d) MS³      (e) SSRAN      (f) DTS

Fig.3 False color image and Error images in MSE and SAM of the city scene AVIRIS dataset.

$$\arg\min_{\mathbf{A}_x}||\mathbf{Y} - \mathbf{LD}_t\mathbf{A}_x||_F^2 + \lambda||\mathbf{G}||_1 + \gamma||\mathbf{H}||_*$$
$$s.t.\ \mathbf{G} = \mathbf{A}_x, \mathbf{H} = [\mathbf{A}_x\ \mathbf{A}_u] \tag{13}$$

Among them, $\mathbf{G} = \mathbf{A}_x$ and $\mathbf{H} = [\mathbf{A}_x\ \mathbf{A}_u]$ denote splitting variables. Then, the closed-form solutions can be calculated by:

$$\mathbf{G} = soft(\mathbf{A}_x - \frac{\mathbf{V}_1}{2\mu}, \frac{\lambda}{2\mu})$$
$$\mathbf{H} = \mathbf{U}soft(\mathbf{\Sigma}, \frac{\gamma}{2\mu})\mathbf{V^T}$$
$$\mathbf{A}_x = ((\mathbf{LD}_t)^\mathbf{T}\mathbf{LD}_t + 2\mu\mathbf{I})^{-1}[(\mathbf{LD}_x)^\mathbf{T}\mathbf{Y} +$$
$$+ \mu(\mathbf{G} + \frac{\mathbf{V_1}}{2\mu}) + \mu(\mathbf{HC} + \frac{\mathbf{V_2C}}{2\mu})] \tag{14}$$

Among them, $\mathbf{V}_1 \in \mathbb{R}^{K \times N}$ and $\mathbf{V}_2 \in \mathbb{R}^{K \times (N+K)}$ denote Lagrangian multiplier, $\mathbf{C} = [\mathbf{I}_{N \times N}\ \mathbf{0}_{N \times K}]^\mathbf{T}$, $\mathbf{U} \in \mathbb{R}^{K \times K}$, $\mathbf{\Sigma} \in \mathbb{R}^{K \times (N+K)}$ and $\mathbf{V} \in \mathbb{R}^{(N+K) \times (N+K)}$ are obtained from the singular value decomposition of $[\mathbf{A}_x\ \mathbf{A}_u] - \mathbf{V}_2/2\mu$.

## III. Experimental Results and Discussions

Relative SOTA spectral super-resolution methods such as, the deep learning based MS³ [10] and SSRAN [11] methods, the dictionary learning based HIRSL [15] and J-SLOL [6] methods, are used in the experiments for better comparison. Besides, two AVIRIS datasets with different scenes are used for comparison, where the false color images of similar scenes and the reconstruction areas are shown in Fig.1.

Table 1 shows the spectral super-resolution performance under four full-reference quality evaluations on the mineral scene. And, the first row of Fig.2 gives the reconstructed results, while the last two rows give the error maps in MSE and SAM. Specifically, the PSNR index of our proposed DTS method is improved more than 1.37dB, and the SAM and ERGAS indexes are decreased by more than 0.09° and 0.11, respectively. The experimental findings presented above show that, our proposed

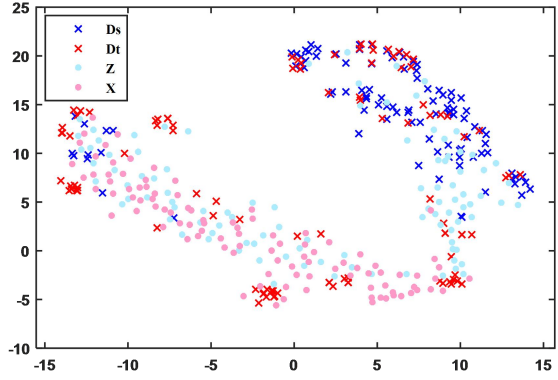

(a) mineral scene

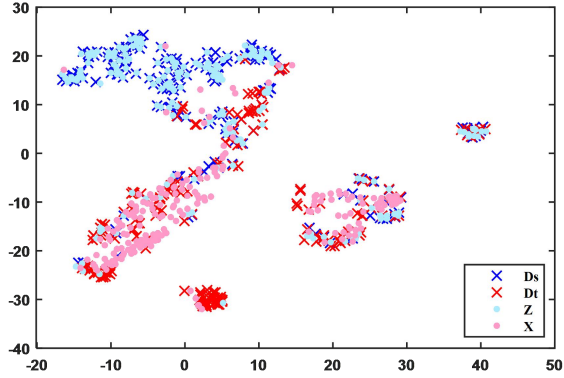

(b) urban scene

Fig.4 Distribution of dictionaries and images before and after dictionary transfer on the AVIRIS datasets corresponding to (a) the mineral scene and (b) the urban scene.

Table 1 Averaged spectral super-resolution results corresponding to the mineral scene AVIRIS dataset.

| Category | Method | MSE | PSNR | SAM | ERGAS |
|---|---|---|---|---|---|
| Deep Learning | MS³ | 0.9023 | 48.5773 | 0.9334 | <u>0.3511</u> |
| | SSRAN | 1.2237 | 47.2542 | 0.8751 | 0.4993 |
| Dictionary Learning | HIRSL | 1.3810 | 46.7289 | 1.2217 | 0.3784 |
| | J-SLOL | <u>0.5887</u> | <u>50.4319</u> | <u>0.7742</u> | 0.3776 |
| | DTS | **0.4287** | **51.8097** | **0.6835** | **0.2395** |

Table 2 Averaged spectral super-resolution results corresponding to the urban scene AVIRIS dataset.

| Category | Method | MSE | PSNR | SAM | ERGAS |
|---|---|---|---|---|---|
| Deep Learning | MS³ | 2.4088 | 44.3127 | 2.8872 | 1.8683 |
| | SSRAN | 1.9733 | 45.1788 | 2.8098 | 2.3301 |
| Dictionary Learning | HIRSL | 2.5531 | 44.0601 | 3.1992 | 1.7126 |
| | J-SLOL | <u>1.7424</u> | <u>45.7194</u> | <u>2.5805</u> | <u>1.5187</u> |
| | DTS | **1.2981** | **46.9978** | **2.4028** | **1.3716** |

DTS method performs better than the other comparison methods both in spatial detail reconstruction and spectral preservation, with the lowest MSE and SAM index. The above results are consistent with the spectral super-resolution performance on the city scene AVIRIS dataset, shown in Table 2 and Fig.3. Among them, the PSNR index of our DTS method is improved more than 1.27dB, and the SAM and ERGAS indexes are decreased by more than 0.17° and 0.14, respectively.

To further visualize and evaluate the effect of dictionary transfer, Fig.4 shows the distribution of dictionaries and images before and after dictionary transfer. As is clearly observable with Fig.4 that, the distributions of $\mathbf{Z}$ and $\mathbf{X}$ are indeed somewhat different. After dictionary transfer, compared with the original spectral dictionary $\mathbf{D}_s$, the distribution of the transferred spectral dictionary $\mathbf{D}_t$ is more consistent with the distribution of $\mathbf{X}$. This phenomenon indicates the significance of the proposed dictionary transfer process, highlighting its indispensability and effectiveness in cross-domain sparse representation.

## IV. CONCLUSION

This paper proposed a compensation matrix based dictionary transfer method for the similar-scene multispectral image spectral super-resolution. The method aims to transfer the spectral dictionary learned in the training domain to the spectral super-resolution domain, and then reconstruct a more accurate HHS image. Specifically, utilizing compensation matrix with similarity constraint, a spectral dictionary transfer scheme is established. Under sparse and low-rank constraints, the corresponding sparse coefficient matrix is optimized. Spectral super-resolution results for AVIRIS datasets show that, our proposed DTS method performs better than the other relative methods.